\documentclass[twocolumn,showpacs,preprintnumbers,amsmath,amssymb]{revtex4}
\usepackage{amsfonts}
\usepackage{amsmath}
\usepackage{graphicx}
\usepackage{dcolumn}
\usepackage{bm}
\usepackage{overpic}
\usepackage{booktabs}

\begin{document}
\title{Reconstructing directed networks for better synchronization}
\author{An Zeng$^{1}$}
\author{Linyuan L\"{u}$^{1}$}\email{linyuan.lue@unifr.ch}
\author{Tao Zhou$^{2}$}

\affiliation{$^1$Department of Physics, University of Fribourg, Chemin du Mus\'{e}e 3, CH-1700 Fribourg, Switzerland\\$^2$Web Sciences Center, University of Electronic Science and Technology of China, Chengdu 611731, People's Republic of China}

\date{\today}

\begin{abstract}
In this paper, we studied the strategies to enhance synchronization on directed networks by manipulating a fixed number of links. We proposed a centrality-based reconstructing (CBR) method, where the node centrality is measured by the well-known PageRank algorithm. Extensive numerical simulation on many modeled networks demonstrated that the CBR method is more effective in facilitating synchronization than the degree-based reconstructing method and random reconstructing method for adding or removing links. The reason is that CBR method can effectively narrow the incoming degree distribution and reinforce the hierarchical structure of the network. Furthermore, we apply the CBR method to links rewiring procedure where at each step one link is removed and one new link is added. The CBR method helps to decide which links should be removed or added. After several steps, the resulted networks are very close to the optimal structure from the evolutionary optimization algorithm. The numerical simulations on the Kuramoto model further demonstrate that our method has advantage in shortening the convergence time to synchronization on directed networks.

\end{abstract}



\maketitle

\section{Introduction}
Synchronization is an important dynamical process in many systems. Understanding and controlling this collective dynamics is of both theoretical and practical significance~\cite{PR93,LiuNature2011}. Many methods have been proposed to enhance the network synchronizability, including the redistribution of the coupling strengths~\cite{PRL034101,PRE016116,PRL218701,PRL138701,ZhaoEPJB2006,LuPRE2007}, the modification of the network structure~\cite{CHAOS037105,CHAOS028101,PRL188701,PRE057102,PRE047102}, the flipping of link directionality~\cite{PRL228702,EPJB217,PNAS10342,ZengR045101}, and so on. Each group of methods has its specific range of applications since different systems are under different technical constrains.

As real systems are frequently manipulated by growing out new connections and eliminating redundant connections~\cite{RMP47}, we here focus on the link reconstructing method to enhance synchronizability. The word ``reconstructing" stands for manipulations including adding, removing and rewiring links, and in each manipulation only one link is created or changed. The reconstructing method for undirected networks has already been studied by using the information embedded in Laplacian eigenvectors~\cite{CHAOS037105}. However, the spectral method cannot be directly applied to directed networks, since the complex value emerges in eigenvectors when the Laplacian matrix is asymmetry. Up to now, how reconstructing methods affect the synchronization on directed networks lacks of systematic consideration.

Previous works show that there are two important factors that mainly affect the synchronization on directed networks: the in-degree distribution~\cite{PRE016116} and the hierarchical structure~\cite{PRE065106,PhysicaD77,EPL48002,NJP043030}. Generally speaking, the more homogeneous the in-degree distribution in a directed network is, the stronger the network synchronizability is. The synchronizability in directed networks can be approximately expressed by $R\approx k_{\texttt{max}}^{\texttt{in}}/k_{\texttt{min}}^{\texttt{in}}$~\cite{PRL034101}. A smaller $R$ indicates a better synchronizability. Accordingly, to enhance synchronizability in directed networks, we can either decrease $k_{\texttt{max}}^{\texttt{in}}$ or increase $k_{\texttt{min}}^{\texttt{in}}$ which are respectively corresponding to removing the links that point to the nodes with maximum in-degree or adding links that point to the nodes with minimum in-degree. Besides the in-degree distribution which plays the dominant role in determining the synchronizability of directed networks, the hierarchical structure may also have considerable effects~\cite{PRE056123}. If the reconstructing strategy only aims at the homogeneous in-degree distribution, an effective hierarchical structure cannot be formed and the resultant synchronizability may fall into local optimum. Therefore, how to choose the starting points of the adding and removing links is very important which may affect the overall hierarchical structure, and thus further affects the synchronizability. In this paper, we employ the PageRank algorithm~\cite{CN1998} to characterize the centralities of nodes on directed networks and design a \emph{centrality-based reconstructing} (CBR) method to enhance synchronizability accordingly. Generally, the information of the hidden hierarchical structure is embedded in the node centrality gradient~\cite{NJP113002}. The nodes with high PageRank scores are more likely to be at the high layer of the network, while those with low PageRank scores probably locate at the low layer. The basic idea of the CBR method is that the newly added links should start from high-layer nodes (with high PageRank scores), while we should choose the links starting form the low-layer nodes (with low PageRank scores) to remove.

Performing the linear stability analysis of synchronizability~\cite{PRL054101,PRL2109,PRE5080} in directed networks, we find that the CBR method is more effective in facilitating synchronization than the \emph{degree-based reconstructing} (DBR) method and \emph{random reconstructing} (RR) method. Furthermore, we apply the CBR method to links rewiring procedure. At each step we first remove one link from the network and then add a new one according to the removing and adding strategies of the CBR method. After several steps, we find that the resulted networks are very close to the optimal structure from the evolutionary optimization algorithm~\cite{EPJB217}. The numerical simulations on the Kuramoto model~\cite{LNP420} further demonstrate that our method has advantage in shortening the convergence time to synchronization on directed networks.

\section{Centrality-Based Reconstructing method}

Consider a directed and unweighted network with $N$ nodes and $E$ links. Denote by $A$ the adjacency matrix where the element $A_{ij}=1$ if there is a directed link from node $i$ to node $j$, otherwise 0. Specifically, $i$ is the start and $j$ is the end. The average degree of the network is $\overline{k}=\frac{E}{N}$. In the synchronization process, each node stands for an oscillator and each directed link represents the influence from one oscillator to another. The synchronization in directed networks is actually formed by a top-down centralized control mechanism~\cite{PRE065106}. The oscillators with higher dynamic centrality are supposed to be more influential in driving the whole system to synchronize. In directed networks, there are many different ways to measure the node's centrality~\cite{CN1998,PRL098701,PRE046114,Hits1999,LeaderRank}. One of the prominent group of methods is based on the random walk process, such as PageRank~\cite{CN1998}, HITs~\cite{Hits1999} and LeaderRank~\cite{LeaderRank} algorithms. In this paper, we apply the well-known PageRank algorithm to quantify the node's centrality in directed network.

PageRank is a famous ranking algorithm which forms the basis of $\rm Google^{TM}$ search engine~\cite{CN1998}. It has been applied to rank scientists in citation networks~\cite{PRE056103} and detect community structure in directed networks~\cite{PRE016103}. In practice, PageRank assigns a score $s_i$ to denote the attractiveness of the webpage $i$. Webpage $i$ obtains higher score if many other important webpages point to it. From the physical perspective, PageRank describes a random walk process on directed network, where the score $s_i$ is proportional to the frequency of visits to a particular node $i$ by a random walker. In PageRank algorithm, a parameter $c$, called \emph{return probability}, is introduced, which represents the probability for a random walker to jump to a random node, and $1-c$ is the probability for the random walker to continue walking through the directed links. Note that, in the oscillator network the link from node $i$ to node $j$ indicates that node $i$ has influence on node $j$, and the nodes who has larger influence on other nodes are more important for the network synchronization. Therefore, in our case the random walker should follow the opposite direction of links. In this way, the node $i$'s centrality score at time $t$ ($t\geq 1$) is given by
\begin{equation}
s_{i}(t)=c+(1-c)\sum \limits^{N}_{j=1}[\frac{A_{ij}}{k_{j}^{in}}(1-\delta_{k_{j}^{in},0})+\frac{1}{N}\delta_{k_{j}^{in},0}]s_{j}(t-1),\label{PR}
\end{equation}
where $\delta_{a,b}=1$ when $a=b$ and $\delta_{a,b}=0$ otherwise. Initially, we assign each node one random walker, namely $s_i(0)=1$ for $i=1,2,\cdots,N$. The typical value of return probability is about 0.15~\cite{CN1998}. The final score of each node is defined as the steady value after the convergence of $s_i(t)$.

Generally speaking, the hierarchical structure of a directed network can be embodied by the node's centrality measured by Eq.~(\ref{PR})~\cite{NJP113002}. A node with higher score is inclined to locate at the high layer of the network hierarchy. With this in mind, we propose a centrality-based reconstructing (CBR) method to enhance the synchronizability on directed networks by manipulating a fixed number of links. Since the network synchronizability can be characterized by the indicator $R\approx k_{\text{max}}^{\texttt{in}}/k_{\texttt{min}}^{\texttt{in}}$: the smaller $R$, the higher synchronizability. Therefore, we have two ways to enhance synchronizability: increasing $k_{\texttt{min}}^{\texttt{in}}$ by adding links pointing to the nodes with minimum in-degree and decreasing $k_{\texttt{max}}^{\texttt{in}}$ by removing links pointing to the nodes with maximum in-degree.
The starting nodes for adding and removing links are selected according to the nodes' centrality scores given by Eq.~(\ref{PR}). To add a link, we choose node $i$ with minimum in-degree as the end, and node $j$ with maximum centrality among all the nodes having not yet pointed to $i$ as the start. To remove a link, we firstly guarantee the end is of maximum in-degree, and then among all possible candidates of starts, we choose the one with minimum centrality. This strategy will help to generate more \emph{receptors}, namely the nodes without any outgoing link, which is a favorable structure for synchronization. For details, one can see the discussion in Sec. III B. Note that, after adding or removing a link the centrality score of each node will be recalculated.

We consider two other methods for comparison: the random reconstructing (RR) method and the degree-based reconstructing (DBR) method. The strategies of choosing the ends of these two methods are the same as CBR, namely adding new link pointing to the node with minimum in-degree and removing link that points to the node with maximum in-degree. The difference is how to choose the starts. When adding or removing links by the RR method, the start of the links are randomly selected. For DBR method, when adding a link, the node with maximum out-degree and not yet pointing to the end will be chosen as the start, while when removing a link, the link starting with a node of minimum out-degree is removed. In the following section, we will compare CBR method with RR and DBR methods on three directed network models based on the linear stability analysis of synchronizability and the numerical simulation on the Kuramoto model.

\section{Results}

\begin{figure}
  \center
  \includegraphics[width=9cm]{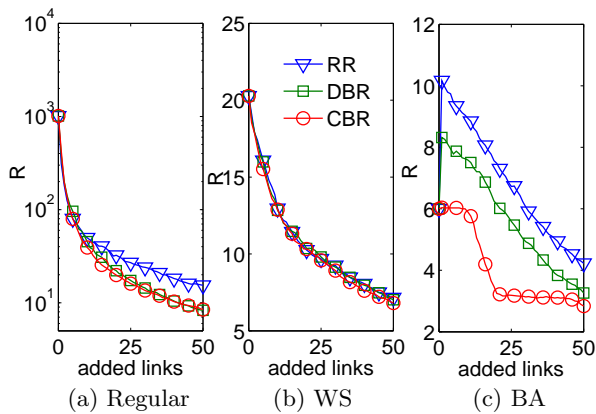}
  \mbox{\hspace{0.3cm}(a) Regular\hspace{1.1cm}(b) WS\hspace{1.6cm}(c) BA}\\
\caption{(Color online) The performance of the RR, DBR and CBR methods on synchronizability when adding links to (a) directed regular networks ($N=100$, $\overline{k}=1$), (b) directed WS networks ($N=100$, $\overline{k}=3$, $p=0.1$) and (c) directed BA networks ($N=100$, $N_{0}=10$, $k_{\text{max}}=6$). The results are averaged over 100 independent realizations.}\label{fig1}
\end{figure}

Under the framework of master stability analysis, the
synchronizability of an undirected network can be quantified by the eigenvalue ratio of
the corresponding Laplacian matrix of this network, namely
$R=\lambda_{N}/\lambda_{2}$, where $\lambda_N$ and $\lambda_2$ are
respectively the largest and the smallest non-zero eigenvalues of
the Laplacian matrix~\cite{PRL054101,PRL2109,PRE5080}. In directed
networks, since the Laplacian matrix, defined as
$L_{ij}=k_{i}^{in}\delta_{ij}-a_{ij}$, is asymmetric with zero
rowsum and has complex eigenvalues. In order to achieve the
synchronization condition, every eigenvalue should be entirely contained in
the region of negative Lyapunov exponent for the particular master
stability function. If the stability zone is bounded and the
imaginary part of complex eigenvalue is small enough, the network
synchronizability can be approximately measured by the real part of
eigenvalue ratio $R=\lambda_{N}^{r}/\lambda_{2}^{r}$ , where
$\lambda_{N}^{r}$ and $\lambda_{2}^{r}$ are respectively the largest
and the second smallest real parts of eigenvalues~\cite{PRE065106,PhysicaD77}. Generally speaking, the smaller the ratio $R$, the stronger
the synchronizability. Here, we employ the indicator $R$ to evaluate the synchronizability
of directed networks.

In this paper, we mainly consider three kinds of directed networks: (i) Directed regular network with identical degree $\overline{k}$ and clockwise links. Here $\overline{k}$ indicates either the average in-degree or the average out-degree since these two values are the same. For example, $\overline{k}=1$ means that each node has one in-link and one out-link. (ii) A variant of Watts-Strogatz (WS) network~\cite{SW} (directed WS network). Starting from a directed regular network, each link will be reconnected with two randomly selected nodes with probability $p\in(0,1)$. (iii) A variant of Barab\'{a}si-Albert (BA) network~\cite{BA} (directed BA network). Starting from a directed tree with $N_{0}$ nodes, a new node with $m$ links ($m$ is a random integer between $1$ and $k_{\text{max}}$) is added to the network in each step until the total node number reaches $N$. Each new added link connects to an existing node $i$ with the probability $q_{i}=\frac{k^{\texttt{in}}_{i}+k^{\texttt{out}}_{i}}{2E}$. The link direction is set to be from older nodes to younger nodes.

\subsection{Adding}

\begin{figure}
  \center
  \includegraphics[width=4.25cm]{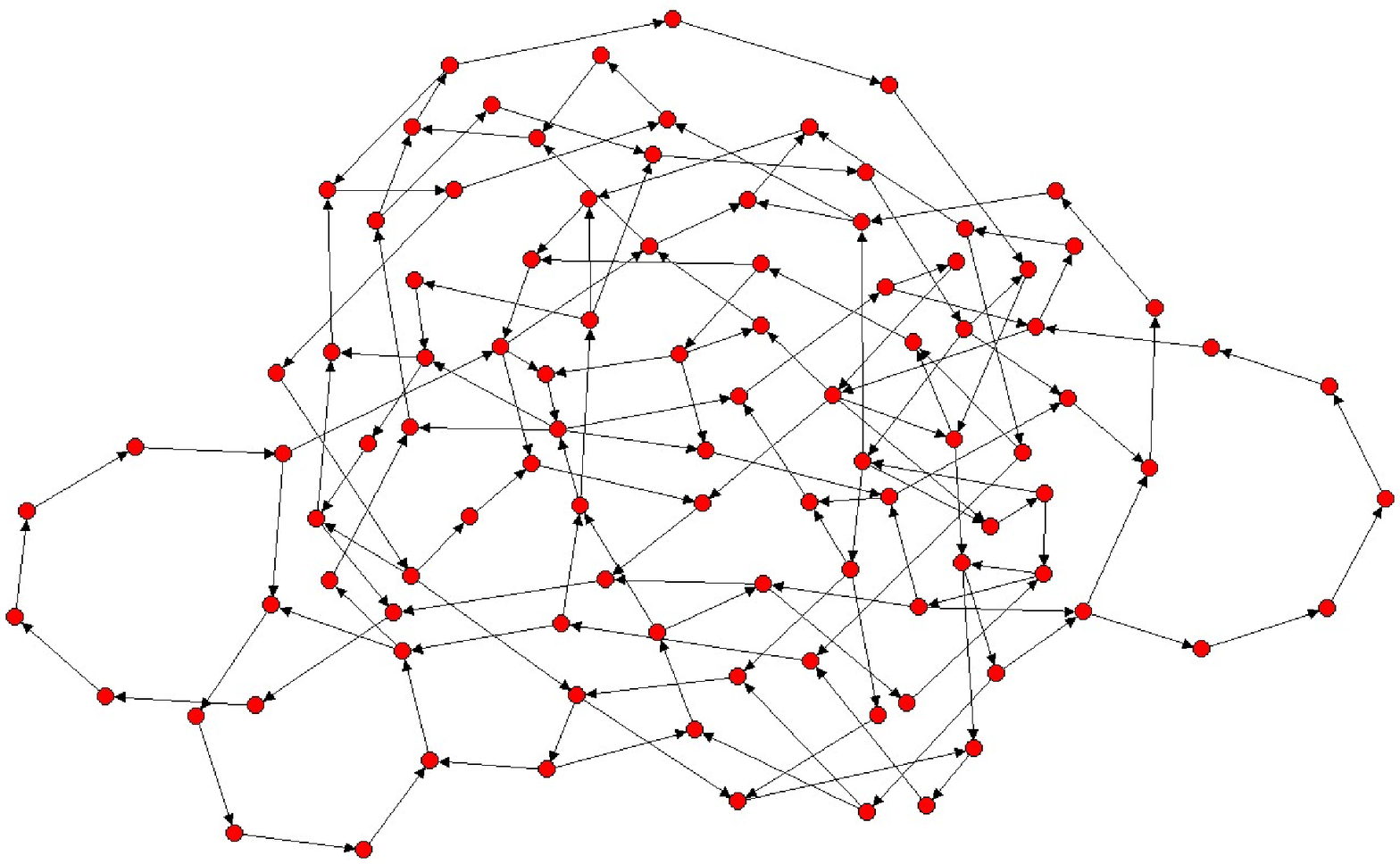}
  \includegraphics[width=4.25cm]{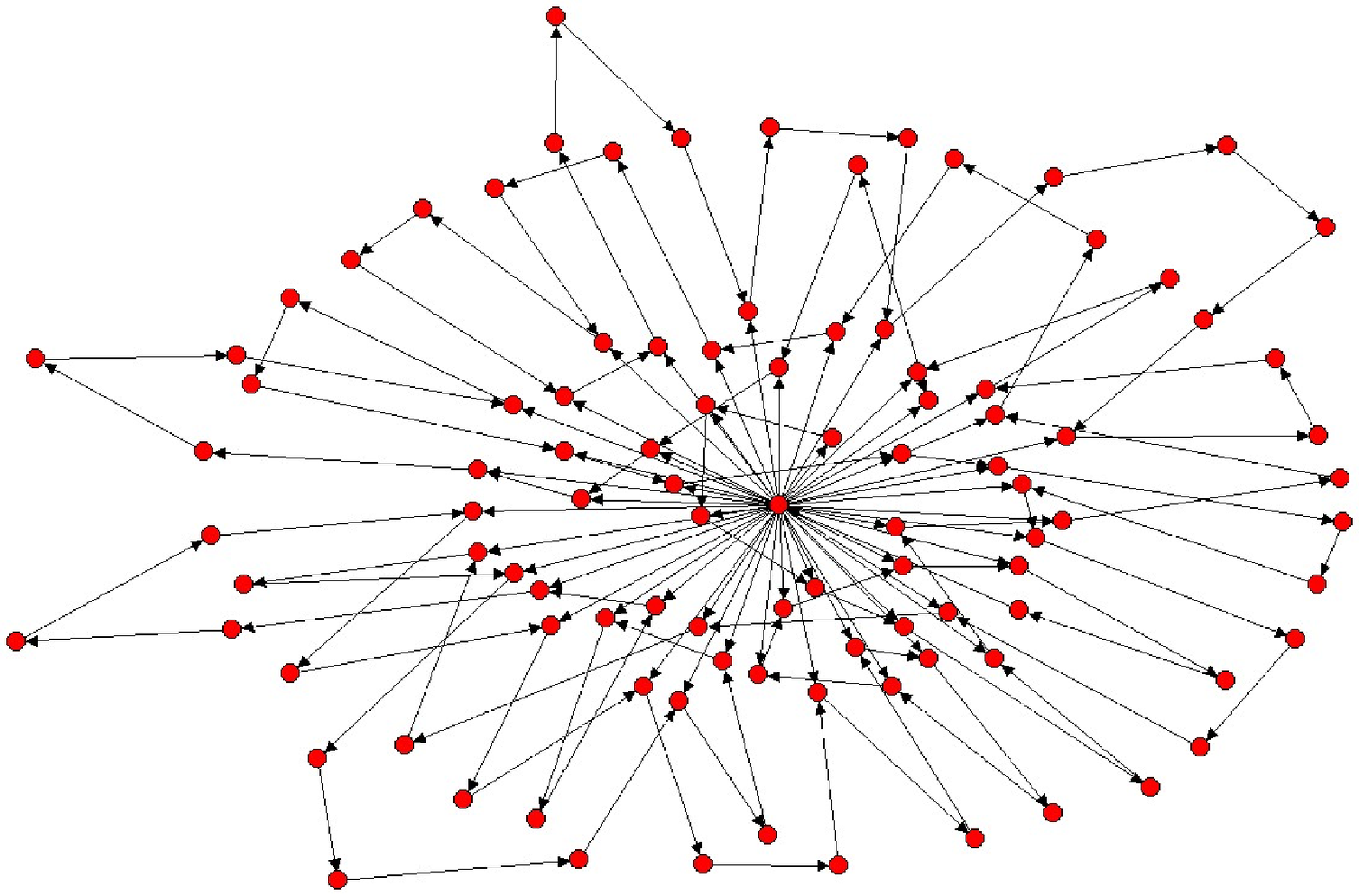}\\
  \mbox{(a)\hspace{4cm}(b)}\\
  \includegraphics[width=4.25cm,height=3cm]{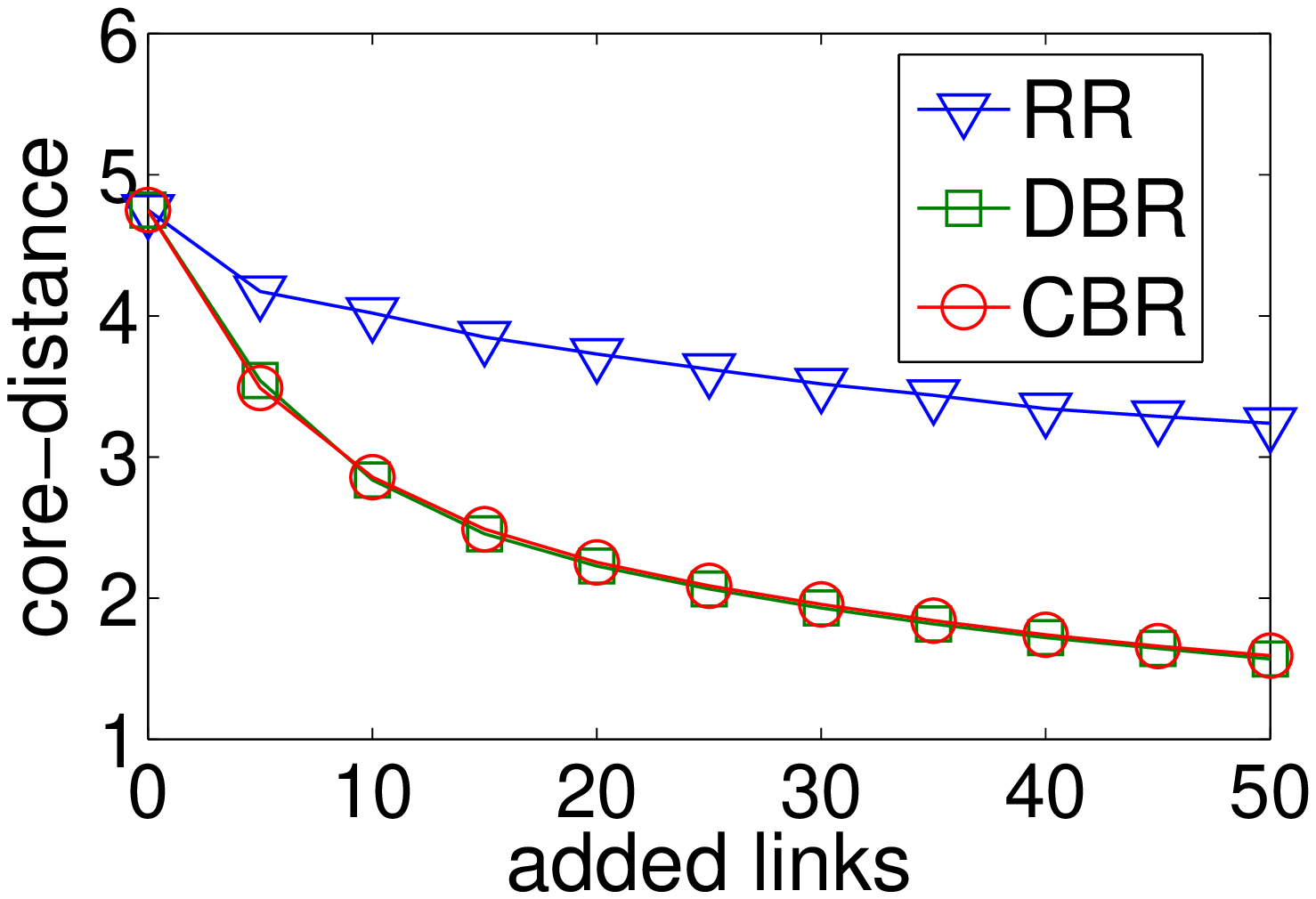}
  \includegraphics[width=4.25cm,height=3cm]{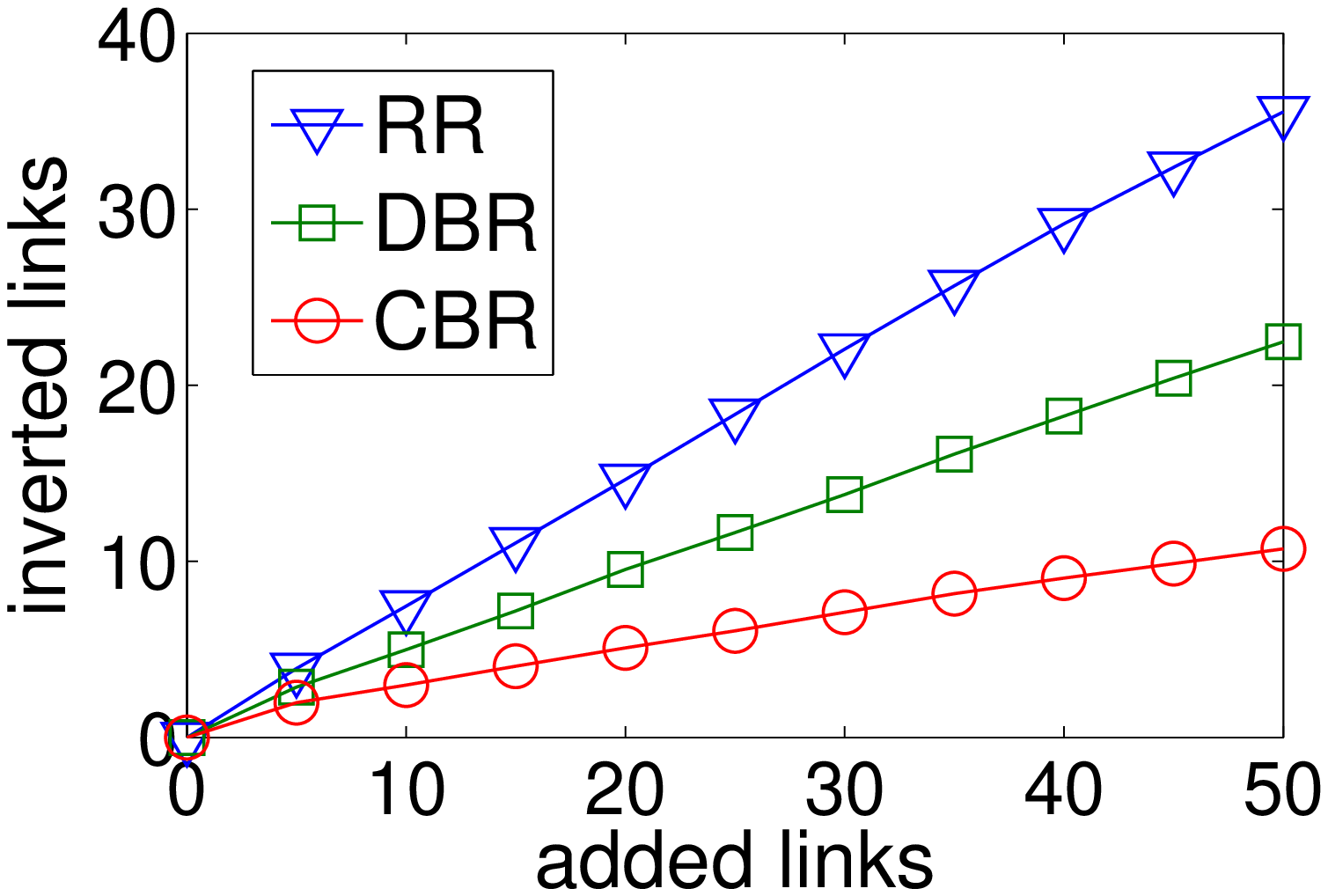}\\
    \mbox{(c)\hspace{4cm}(d)}\\
\caption{(Color online) (a) The obtained network by adding $50$ links to a directed regular network with the RR method. (b) The obtained network by adding $50$ links to a directed regular network with the CBR method. (c) The dependence of core-distance on the number of the added links in the directed WS networks. (d) The dependence of the number of inverted links on the number of the added links in the directed BA networks. The parameters of the network models are the same as those in Fig.~\ref{fig1}. The results in (c) and (d) are averaged over 100 independent realizations.}\label{fig2}
\end{figure}

The results for adding links in three modeled networks are reported in
Fig. 1. Generally speaking, CBR performs best among all three
methods. Specifically, in directed regular networks [see Fig. 1(a)] all these three methods can decrease $R$ by adding links.
The enhancement of synchronizability results from the decreasing of the average shortest distance of networks. As we know, networks with small average shortest distance
are likely to have better synchronizability~\cite{Hong2004,ZhaoPhysicaA2006}. Besides,
the DBR method and CBR method outperform the RR
method especially when many links are added. The reason is that the DBR method and CBR method are capable to generate a network core during the adding process, which
may involve one (when adding a small number of links) or several
(when adding many links) nodes. The core node has many outgoing links and drives other nodes during the synchronizing process. It has been pointed out that such a core is favorable for the network
synchronizability~\cite{EPJB217}. In contrast, the RR method is unable to induce such an effective core, thus its synchronizability enhancement is only from the decreasing of the average shortest distance. Two
typical examples are shown in Fig. 2(a) and Fig. 2(b) which are obtained by adding $50$ links to a directed regular networks with RR and CBR methods, respectively. Clearly, there exists a center node
as the core of the obtained CBR network, yet there is no observable central nodes for the obtained RR network.

Figure~\ref{fig1}(b) shows that the indicator $R$ decreases with the increasing number
of added links for all these three methods and the difference between them is very small ($R_{\texttt{CBR}}=6.81$,$R_{\texttt{DBR}}=7.02$,$R_{\texttt{RR}}=7.16$). The enhancement mainly comes
from narrowing the in-degree distribution. For the directed WS networks, even though the CBR and DBR methods can still form a core by adding a few links, the positive effect of the core is disrupted by the
complicated structure of the initial network. Therefore, when adding a few links, like 50, comparing with RR method the synchronizability enhancements by CBR and DBR methods are very limited, respectively 4.9\% and 2.0\%. Note that, when we add more links, like 500, the improvements increase to 12.3\% and 9.7\% for the CBR and DBR methods ($R_{\texttt{CBR}}=1.71$,$R_{\texttt{DBR}}=1.76$,$R_{\texttt{RR}}=1.95$), respectively, since the effect of the initial network will be depressed. Additionally, the CBR and DBR methods can lead to shorter convergence time than the RR method due to the smaller core-distance (see details in Sec. III D for Kuramoto model). Given a network, the core-distance is defined as the average shortest distance from the core node to all other nodes in this network. It was pointed out that the smaller the core-distance, the quicker the whole network converges to synchronization, since the core actually drives the whole network to the synchronized state~\cite{EPJB217}. The dependence of the core-distance on the number of added links in the directed WS networks is shown in Fig.~\ref{fig2}(c). Note that, since there is no core in RR network, we use the minimal distance from a node to all other nodes in the network (such node is considered as a core). Clearly, the values of the core-distance of the CBR and DBR methods are much smaller than that of RR, and thus the CBR and DBR methods can result in shorter convergence time.

In directed BA models, as shown in Fig.~\ref{fig1}(c) it is clear that CBR method performs best. The directed BA model has obvious hierarchical
structure where the nodes with higher centrality scores incline to locate
at the higher layer. In CBR method,
the start of the new link is chosen as the node with
maximum centrality score, thus the generated core is actually the
root node (in the highest layer) of the initial directed BA
networks. In contrast, with DBR method the core nodes
usually locate in low layers. With starts in low layers many inverse links (i.e., links from low layers to high layers) will be formed, leading to a large number of loops. The
situation is even worse with RR method. The dependence of the
number of inverse links on the number of added links of BA networks is
shown in Fig.~\ref{fig2}(d). We can see that the number of inverse links generated by
CBR method is fewer than DBR method and RR method. The large number of loops are not
beneficial to synchronization~\cite{PRE065106,EPJB89}, and thus in directed BA networks CBR method can
prominently enhance the network synchronizability than the other two
methods. The suddenly jump of the indicator $R$ for RR and DBR methods after adding one link is coursed by the existent of a long loop that contains the root node. This loop breaks the leadership of root node in driving the synchronizing process.

\subsection{Removing}

\begin{figure}
  \center
  \includegraphics[width=9cm]{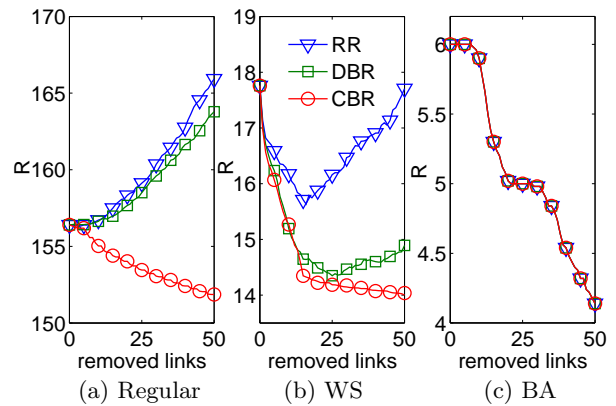}
  \mbox{\hspace{0.3cm}(a) Regular\hspace{1.1cm}(b) WS\hspace{1.6cm}(c) BA}\\
\caption{(Color online) The performance of the RR, DBR and CBR methods on synchronizability when removing links from (a) directed regular networks ($N=100$, $\overline{k}=3$), (b) directed WS networks ($N=100$, $\overline{k}=3$, $p=0.1$) and (c) directed BA networks ($N=100$, $N_{0}=10$, $k_{\text{max}}=6$). The results are averaged over 100 independent realizations.}\label{fig3}
\end{figure}

\begin{figure}
  \center
  \includegraphics[width=8.5cm]{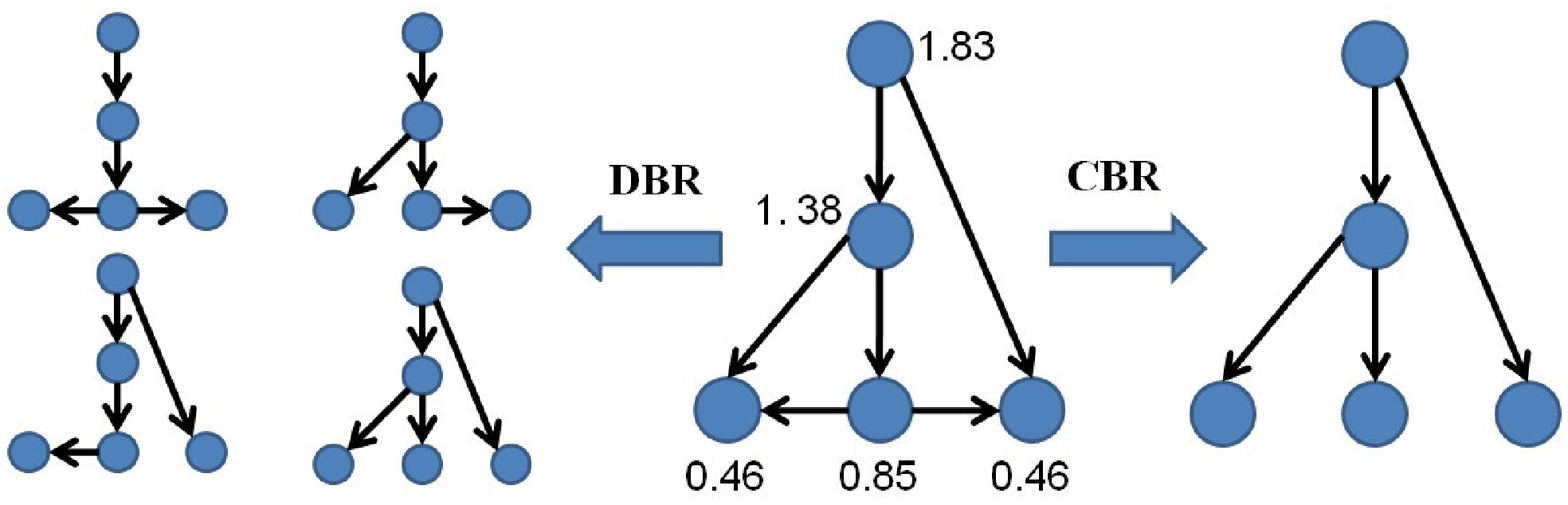}\\
  \mbox{(a)}\\
  \includegraphics[width=4.25cm,height=3cm]{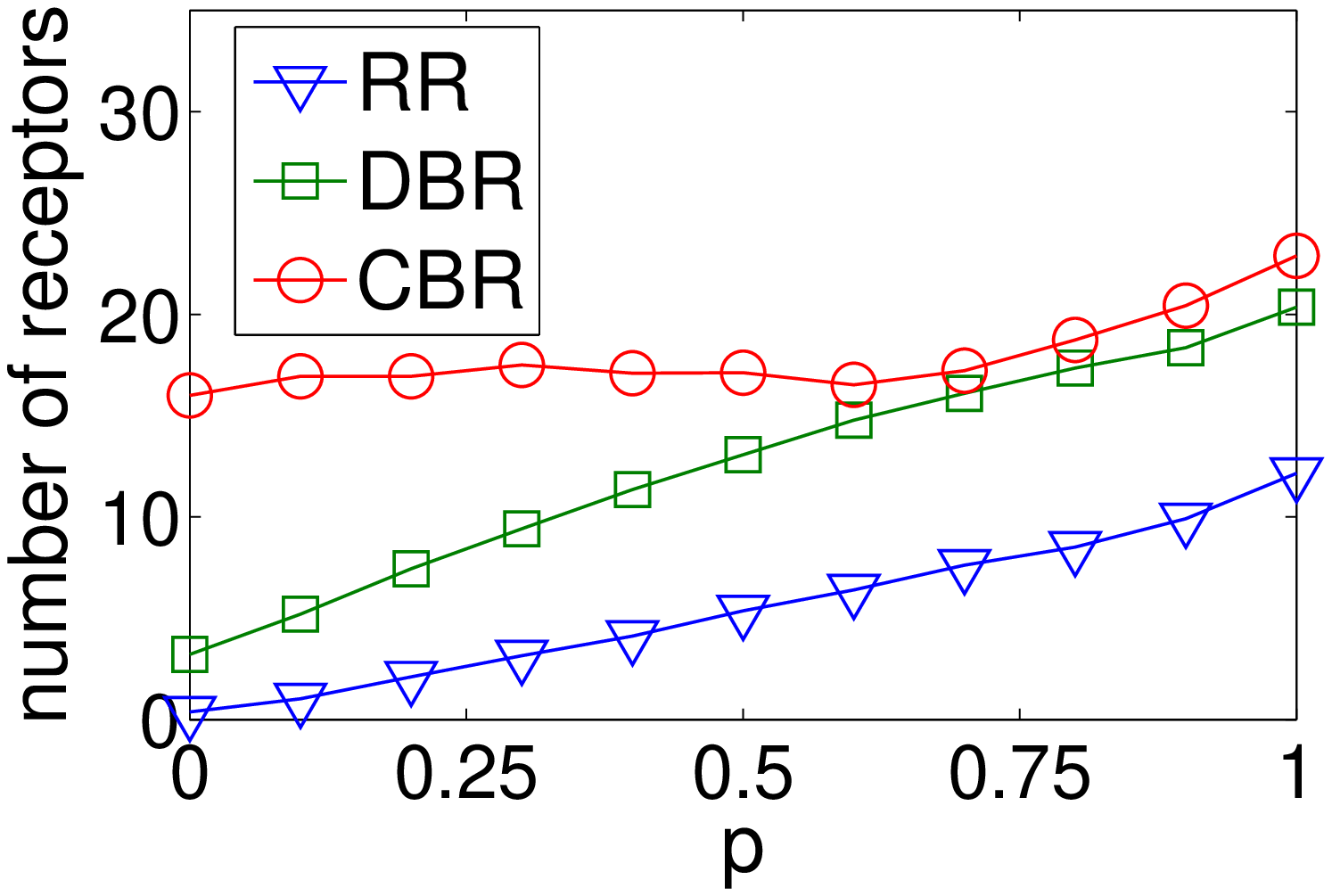}
  \includegraphics[width=4.25cm,height=3cm]{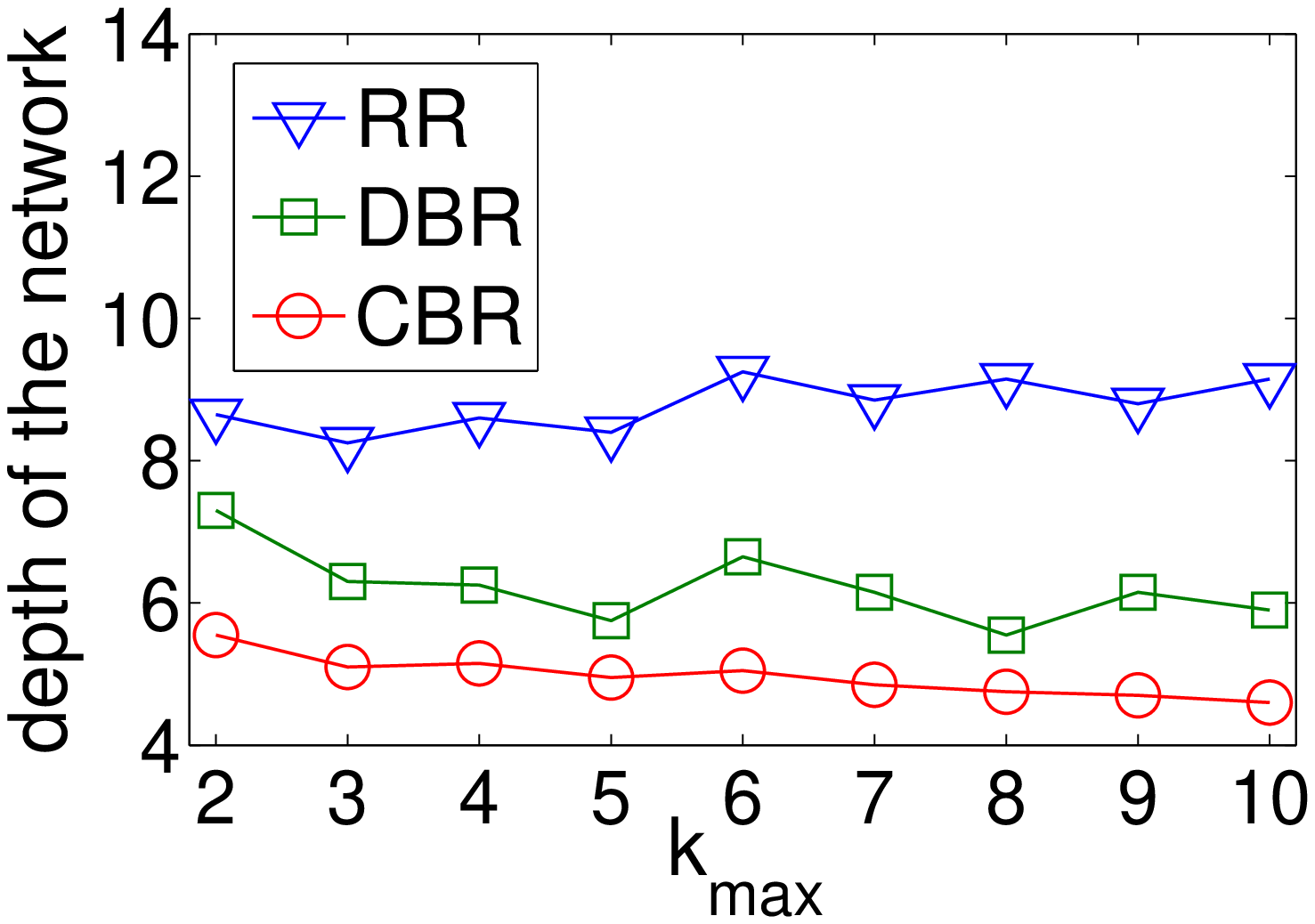}\\
    \mbox{(b)\hspace{4cm}(c)}\\
\caption{(Color online) (a) A simple example to illustrate why CBR method has the advantage in shortening the depth after removing links from directed BA networks. The nodes' centrality scores of the original network are labeled beside the nodes. (b) The number of generated receptors when cutting 50 links from directed WS networks by three methods. The original networks are given $N=100$ and $\overline{k}=3$ but with different $p$. (c) The depth of the obtained trees by three methods from directed BA networks. The original networks are given $N=100$ and $N_{0}=10$ but with different $k_{\text{max}}$.  The results in (b) and (c) are averaged over 100 independent realizations.}\label{fig4}
\end{figure}

When removing links, the CBR method can enhance the
synchronizability more effectively than the other two methods as
shown in Fig.~\ref{fig3}. In directed
regular networks [see Fig.~\ref{fig3}(a)], both the RR method and the DBR method
weaken the synchronizability while the CBR method
improves the synchronizability. By removing links, all three methods will destroy the uniform in-degree
distribution in directed regular networks. However, CBR method can generate lots of receptors (i.e., the nodes without any outgoing link) who are only affected by other nodes. This
structure is favorable for synchronization since these receptors do not send back interruptive information to their upstream nodes. When removing the first link from the regular network, since the out-degree and the centrality score for all the nodes are the same, it is equivalent to a random selection of one link. After that, the node whose out link was deleted will have the lowest centrality score. Therefore, in the next step a link starting from this node will again be removed, until all its out-links were removed. As a result, this node becomes a receptor. Figure ~\ref{fig4}(b) shows the number of generated receptors of the obtained networks when cutting 50 links by three methods, respectively. The parameter $p=0$ corresponds to regular networks, $p=1$ to random networks, and $0<p<1$ to WS networks. Clearly, CBR method can generate more receptors than RR and DBR methods.

In directed WS networks, the CBR method also performs the best as shown in
Fig.~\ref{fig3}(b). At the beginning, three methods can
enhance the synchronizability by reducing the maximum in-degree in
the networks. After the in-degree distribution becomes homogeneous, the situation will become similar to the directed regular networks. If we go on removing links by RR method or DBR method, the synchronizability will be weaken
since the homogeneous in-degree distributions are broken. By creating
many receptors, CBR method can further enhance the synchronizability
even the homogeneous in-degree distribution gets destroyed.

In directed BA models, the synchronizability is identical in three
methods since in acyclic networks the synchronizability is totally determined by the
maximum and minimum in-degree. In Fig.~\ref{fig3}(c),
each stair is corresponding to a typical maximum in-degree of the
networks. Even though the three methods obtain the same
synchronizability, the convergence time to synchronization are
different. By removing the links from nodes in relatively lower layers, the obtained networks will have shorter depth, and thus converge faster~\cite{EPL48002,NJP043030}. Figure~\ref{fig4}(a) gives a typical example to compare DBR and CBR methods. We can see that after removing two links by the CBR method from the middle plot the \emph{depth} of the obtained tree is 2. While the DBR method can lead to four different trees with equal probability. Since three of them are of depth 3 and the rest is of depth 2, the expected depth of the obtained tree by DBR is 2.75. The statistical results on the depth of the obtained trees after removing links from the directed BA models are shown in Fig.~\ref{fig4}(c). Clearly, the CBR method can generate a tree with shorter depth than the RR and DBR methods, and thus the networks obtained by the CBR method can converge faster to the synchronized state. See the Kuromoto model in Sec. III D for numerical evidence.

\subsection{Rewiring}

An integrated way is to enhance the synchronizability
by rewiring links while keeping the total number
of links unchanged. In each step, we firstly remove
a link from the network according to the strategy of removing links (see
Sec. III B) and then add a link according to the strategy of adding links
(see Sec. III A). Starting from different network models, we rewire $300$ links via the RR, DBR and CBR methods, respectively. The synchronizability $R$ of the obtained networks are shown in Table ~\ref{table}. The corresponding optimal synchronizability for such directed networks is obtained by an evolutionary optimization algorithm~\cite{EPJB217,PNAS10342}. We find that the synchronizability corresponding to the CBR method is very close to
the optimal case. In addition, the
topology of the obtained CBR network  is
very similar to the optimal case, such as the hierarchical structures and the existence of cores. These
results indicate that the CBR method is an effective method to enhance the synchronizability via link rewiring.

\begin{table}[!htb]
 \extrarowheight=0.4em
 \tabcolsep=4pt
\begin{center}
\caption{The synchronizability of the networks obtained by rewiring 300 links through RR, DBR and CBR methods. The network parameters are set as follows: $N=100$ and $\overline{k}=3$ for regular networks; $N=100$, $\overline{k}=3$ and $p=0.1$ for WS networks; $N=100$, $\overline{k}=3$ and $p=1$ for Erd\"os-R\'{e}nyi (ER) networks~\cite{ER}; $N=100$, $N_0=10$, $k_{\text{max}}=6$ and $\overline{k}=3$ for BA networks. The sychronizability of the original networks ($R_0$) and the result obtained by the evolutionary optimization algorithm ($R_{\texttt{opt}}$) for the network with $N=100$ and $\overline{k}=3$ ~\cite{EPJB217} are shown for comparison.}
\begin{tabular}{lccclccc}
\hline
  network  &Regular &WS  &ER &BA\\
\hline
  $R_{0}$ &$156.43$ &$18.384$ & $239.49$ & $6.0000$\\
  $R_{\texttt{RR}}$  &$3.6975$ &$3.4459$ & $3.4043$ & $3.2231$\\
  $R_{\texttt{DBR}}$  &$1.3813$ &$1.4511$ & $1.4415$ & $1.3822$\\
  $R_{\texttt{CBR}}$ &$1.2191$ &$1.2694$ & $ 1.3401$ & $1.3333$\\
  $R_{\texttt{opt}}$  &$1.17$ &$1.17$ & $1.17$ & $1.17$\\
 \hline\label{table}
\end{tabular}
\end{center}
\end{table}

\subsection{Kuramoto model}

\begin{figure}
  \center
  \includegraphics[width=4.25cm]{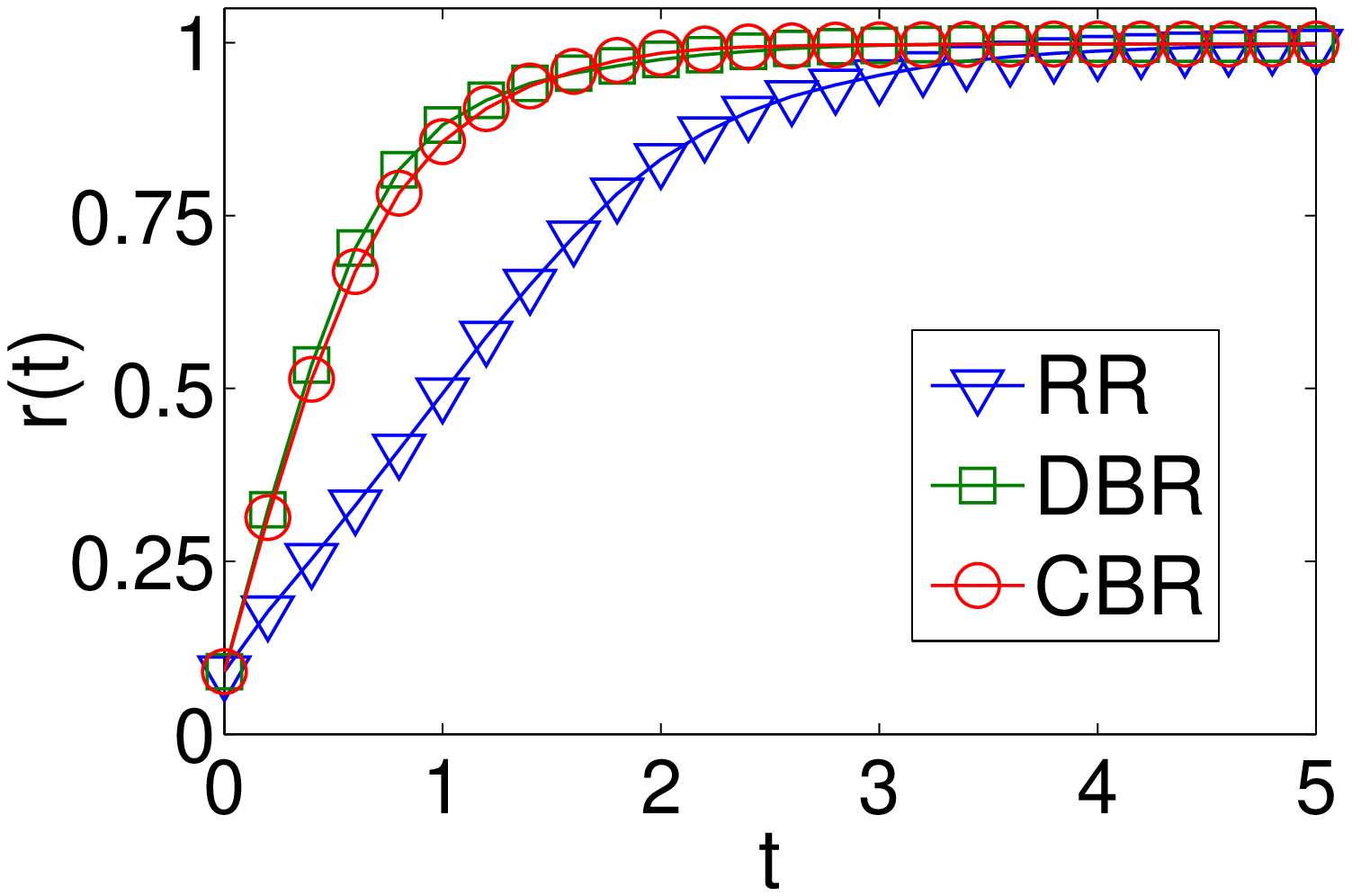}
  \includegraphics[width=4.25cm]{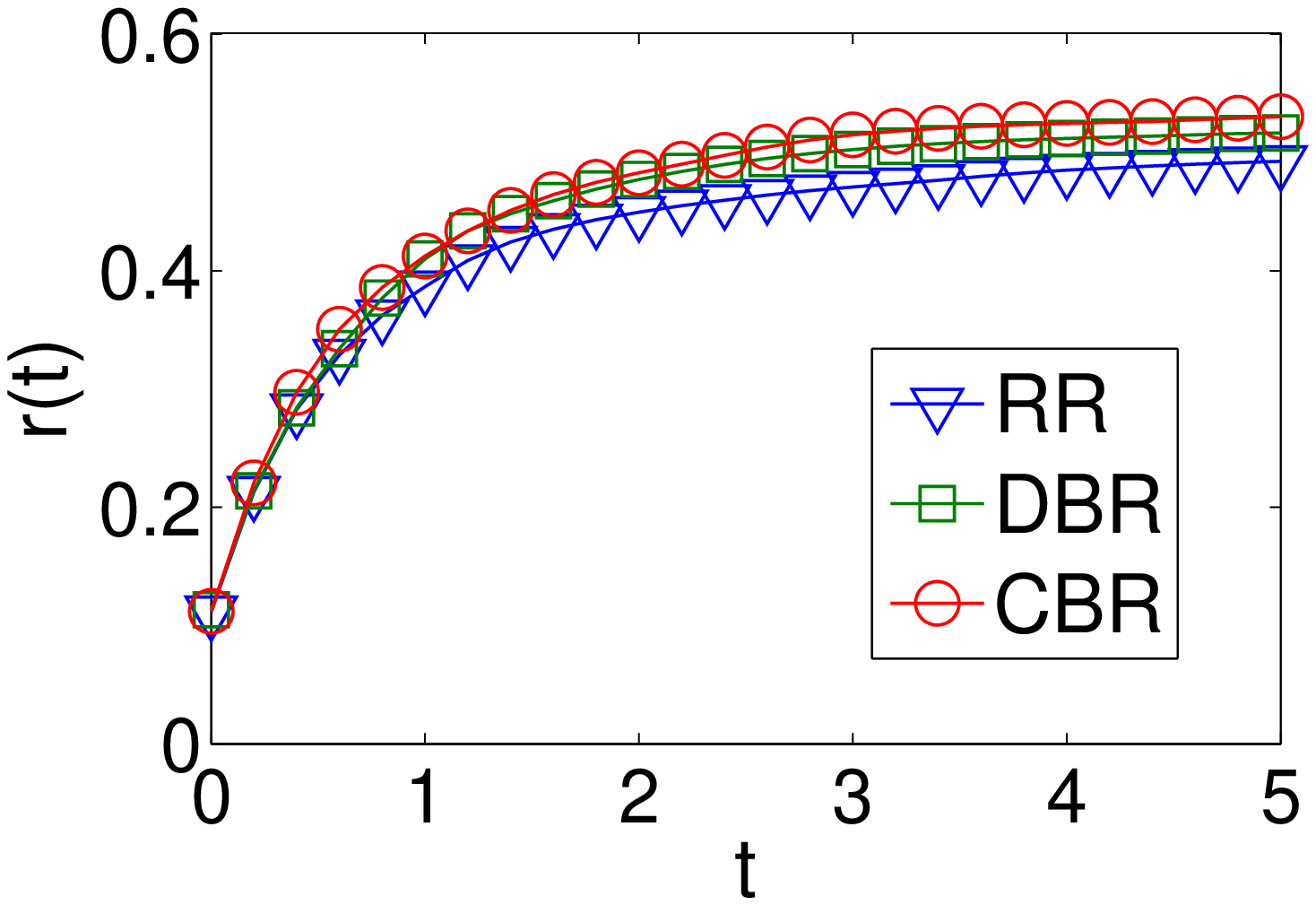}\\
    \mbox{(a) Regular\hspace{3cm}(b) Regular}\\
  \includegraphics[width=4.25cm]{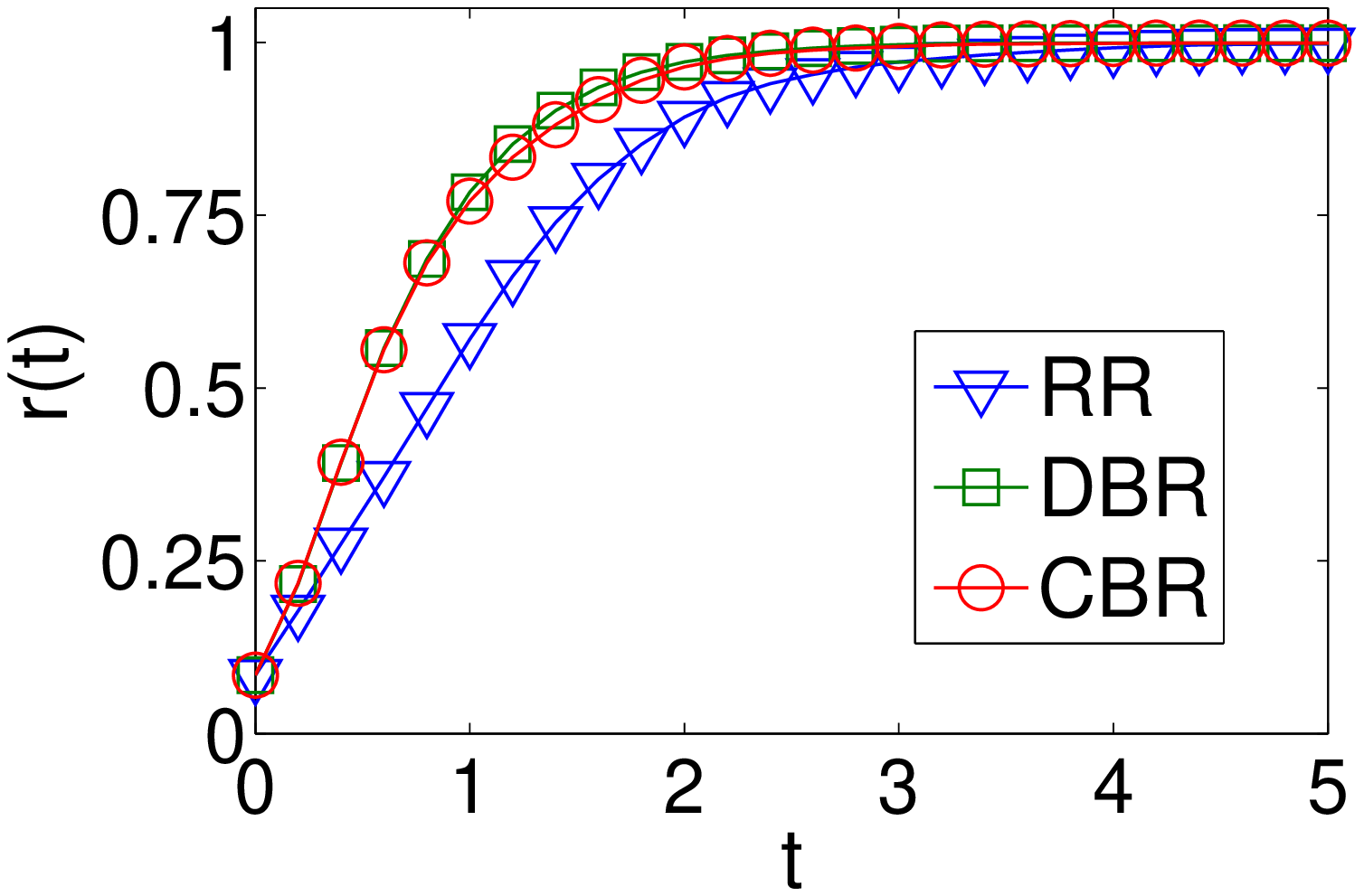}
  \includegraphics[width=4.25cm]{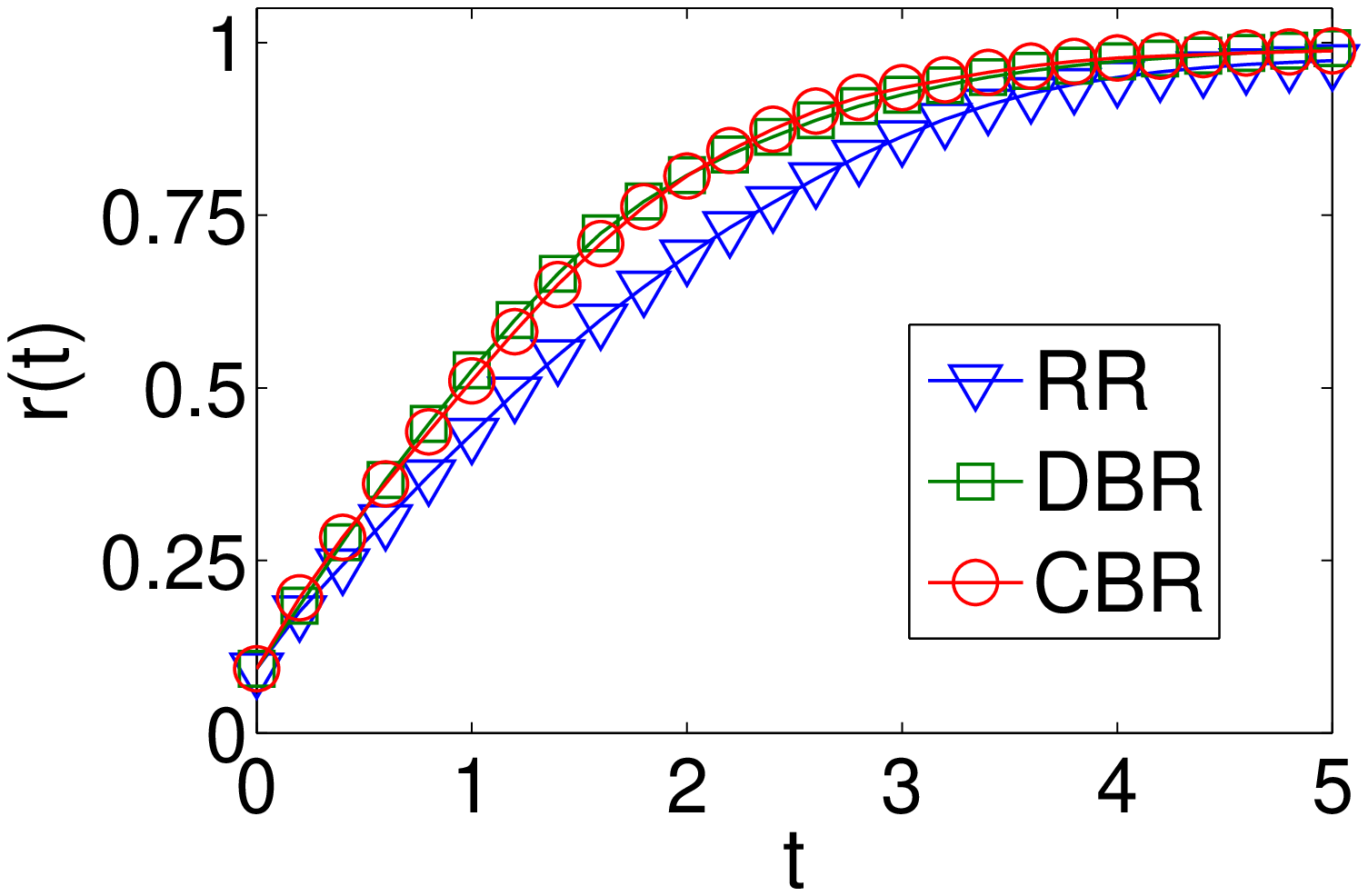}\\
    \mbox{(c) WS\hspace{3.5cm}(d) WS}\\
  \includegraphics[width=4.25cm]{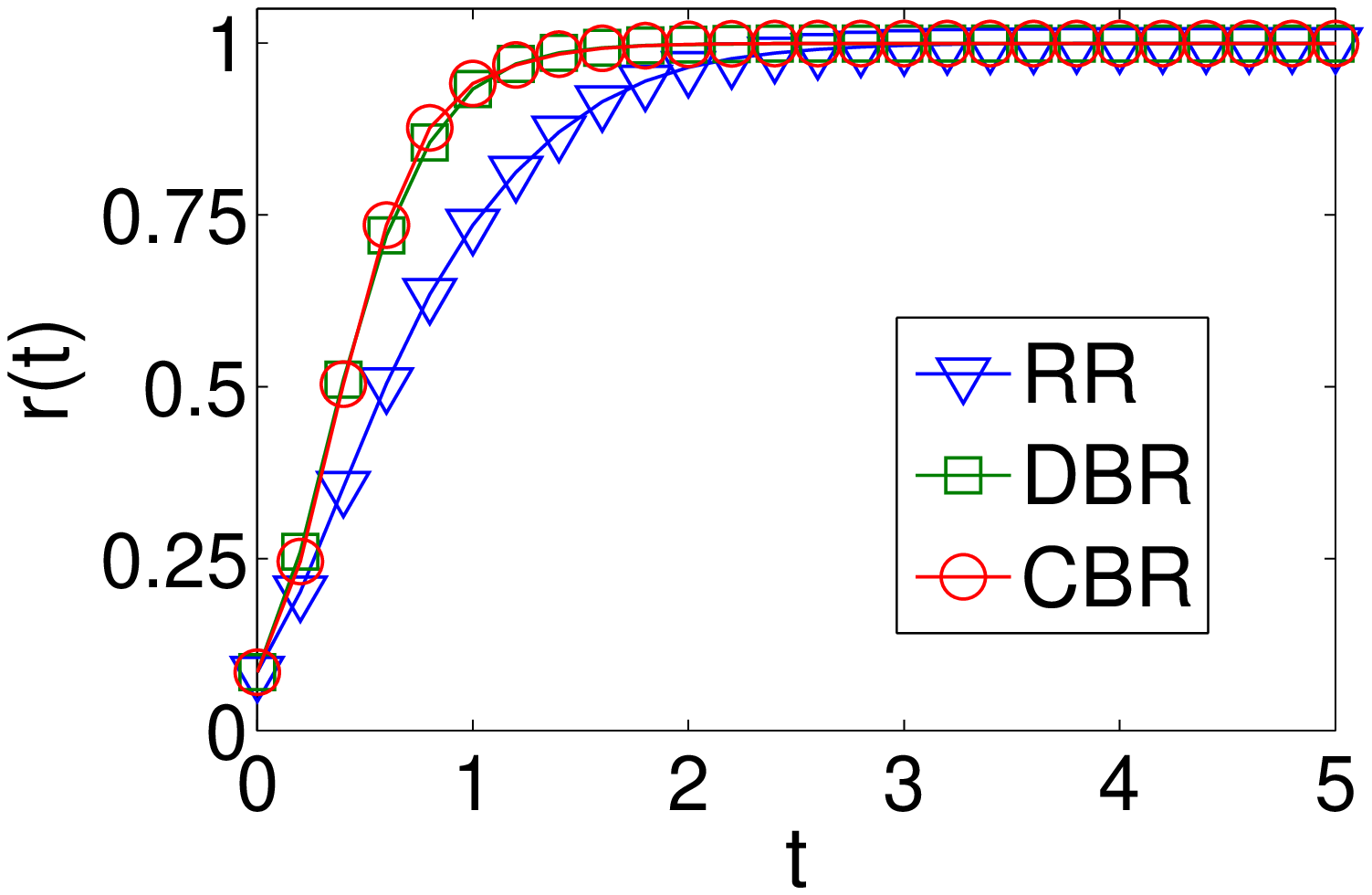}
  \includegraphics[width=4.25cm]{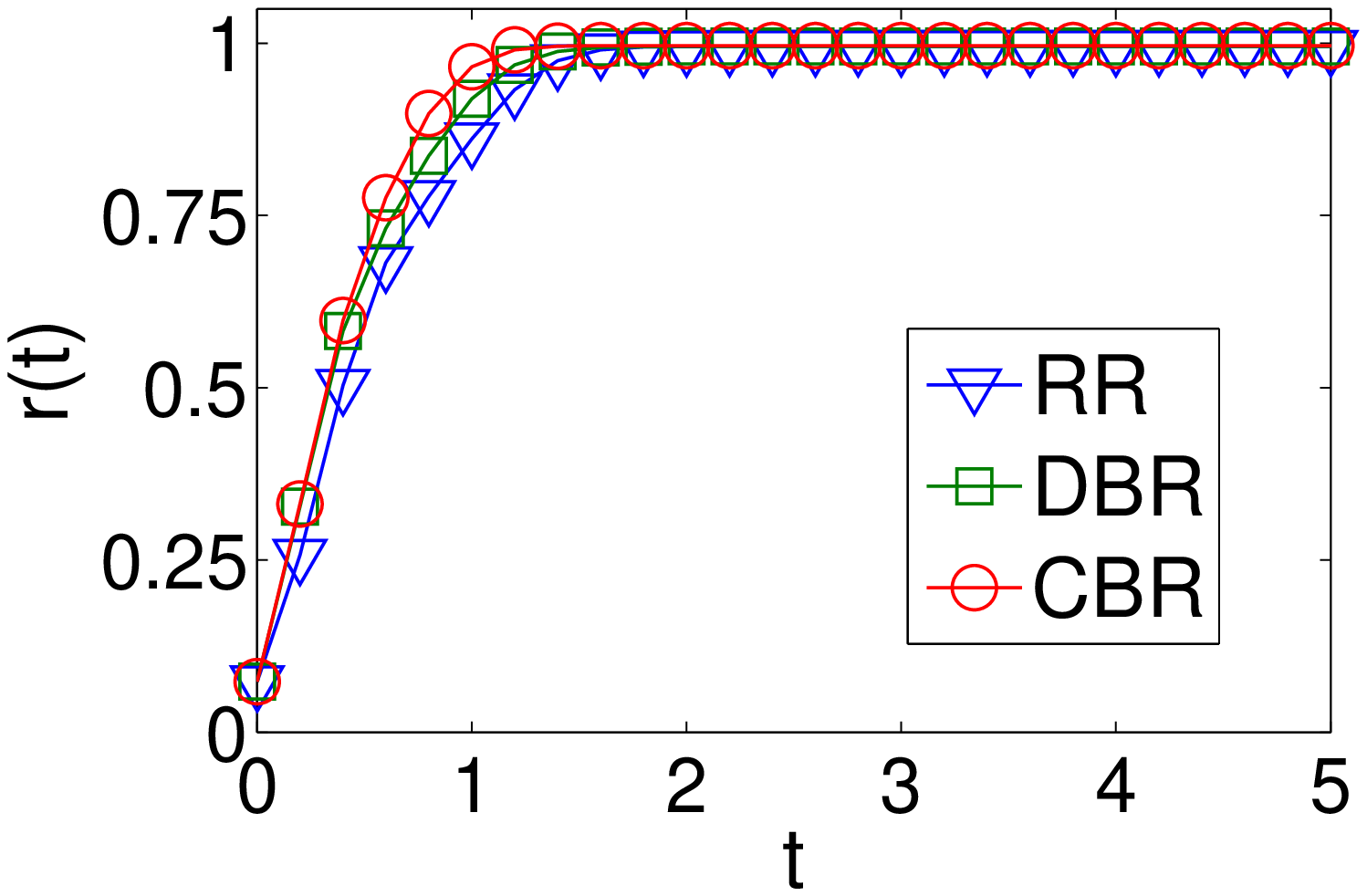}\\
     \mbox{(e) BA\hspace{3.5cm}(f) BA}\\
\caption{(Color online) The order parameter of Kuramoto model on (a)
obtained networks by adding $50$ links to directed regular
networks, (b) obtained networks by removing $50$ links from directed
regular networks, (c) obtained networks by adding $50$ links to
directed WS networks, (d) obtained networks by removing $50$ links
from directed WS networks, (e) obtained networks by adding $50$
links to directed BA networks, (f) obtained trees by removing links from directed BA networks. The network parameters are the same as in Fig.~\ref{fig1}. The results are averaged over 100
implements.}\label{fig5}
\end{figure}

For the purpose of monitoring the effects of different rewiring methods on the convergence time to synchronization, we
study the Kuramoto model~\cite{LNP420} in the resulted networks. The
dynamics of $N$ coupled oscillators is described by the equation
\begin{equation}
\dot{\theta_{i}}=\omega_{i}+\frac{K}{\overline{k}}\sum_{j=1}^{N}
A_{ji}\mathrm{sin}(\theta_{j}-\theta_{i}),
\end{equation}
where $\omega_{i}$ and $\theta_{i}$ are the natural frequency and
the phrase of oscillator $i$, respectively. The coupling strength $K$ is set to be a
positive constant, here we fix $K=10$ for convenience. The
collective dynamics of the whole population is measured by the
macroscopic complex order parameter
\begin{equation}
r(t)e^{i\phi(t)}=\frac{1}{N}\sum_{j=1}^{N}e^{i\theta_{j}(t)},
\end{equation}
where the modulus $0\leq r(t) \leq 1$ measures the phase coherence
of the population and $\phi(t)$ is the average phase. $r(t)\simeq1$
and $r(t)\simeq0$ describe the limits in which all oscillators are respectively phase locked and moving incoherently.

In Fig.~\ref{fig5}, we compare the behavior of Kuramoto model on the obtained
networks from (a) adding 50 links to directed regular networks, (b)
removing 50 links from directed regular networks, (c) adding 50 links
to directed WS networks, (d) removing 50 links from directed WS
model, (e) adding 50 links to directed BA networks and (f) obtained
tree by removing links from directed BA networks.

Generally speaking, the CBR and DBR methods outperform the RR method in the convergence time to the completely
synchronized state. As mentioned above, the emergence of the network core can greatly shorten the convergence time. The smaller the core-distance, the quicker the whole network converges to synchronization, since the core actually drives the whole network to the synchronized state. In Fig.~\ref{fig5}(a), (c) and (e), it is obvious that the CBR and DBR networks have shorter convergence time than the RR method due to the smaller core-distance.

In Fig.~\ref{fig5}(b), the network cannot reach the complete synchronization. However, it can be seen that the order parameter of CBR networks is higher than the RR and DBR networks, in accordance with the higher synchronizability predicted by the master stability analysis. Figure~\ref{fig5}(d) shows that the networks from the CBR and DBR methods converge faster than the ones from the RR method. It is because RR is more likely to remove long range links, resulting in an increasing of the average distance of the network, which usually corresponds to a longer convergence time. In contrast, in the directed WS networks, the starts of the long range links usually have relatively large out-degrees and high PageRank scores, and therefore are not likely to be removed according to the CBR or DBR methods. As shown in Fig.~\ref{fig5}(f), the CBR method leads to the smallest convergence time among the three methods. As the PageRank centrality score can help the CBR method to recognize the hierarchical structure of the directed BA model, the final spanning trees of the CBR method are always with fewer depth than the DBR and RR methods [see also Fig.~\ref{fig4}(c)] resulting in the faster convergence.

\section{Conclusion}

In summary, we proposed a centrality-based reconstructing (CBR) method to facilitate synchronization in
directed networks. The centrality is measured by the PageRank algorithm. Numerical analysis on three network models shows that the CBR method is more
effective in enhancing the network synchronizability than the degree-based reconstructing method and
random reconstructing method. Specifically, when adding links, the
CBR method forms a wisely located core to drive the whole network
to the synchronized state. When removing links, the CBR method generates
lots of receptors which only receive information and do not
disturb the upstream nodes. Furthermore,
the CBR method can also be extended to deal with the link rewiring problem. Significantly, the obtained
network through CBR link rewiring method has very close synchronizability to the
corresponding optimal case obtained by the evolutionary
optimization algorithm. Finally, the result from the Kuramoto model shows that the CBR method has advantage in shortening the convergence time to synchronization.

In practice, as real systems are frequently manipulated by
growing new connections and eliminating redundant
connections~\cite{RMP47}, the CBR method can provide a useful way to
improve synchronizability. In some specific systems like
ecosystems~\cite{SCI1360} and neuron systems~\cite{NRN285},
synchronization should be inhibited. In such situation, the inverted
CBR method can be applied to decrease the synchronizability, namely to add links starting from the nodes with low centrality scores to the nodes with high in-degrees and to remove the links whose starts have high centrality scores while whose ends own low in-degrees.

\section*{Acknowledgement}
This work is supported by the Swiss National Science Foundation
under Grant No. 200020-132253.

\end{document}